\documentclass[runningheads]{llncs}
\usepackage{graphicx}
\usepackage[misc]{ifsym}
\usepackage{amsmath}
\usepackage{amssymb}

\usepackage{bbding}
\usepackage{pifont}
\usepackage{wasysym}
\usepackage{utfsym}
\usepackage{fontawesome}
\usepackage{url}
\usepackage{subcaption}
\usepackage{mathtools}
\usepackage{color}
\usepackage{multirow}
\usepackage{algorithm}
\usepackage{algorithmic}
\usepackage{amsmath}
\usepackage{gensymb}
\usepackage{booktabs}
\usepackage{cite}
\usepackage{xcolor}

\newcommand{\repeatthanks}{\textsuperscript{\thefootnote}}
\newcommand{\methodname}{DDAug}

\usepackage{hyperref}
\hypersetup{hidelinks,
	colorlinks=true,
	allcolors=blue,
	pdfstartview=Fit,
	breaklinks=true}

\begin{document}
%
% \title{Contribution Title\thanks{Supported by organization x.}}

\title{Dynamic Data Augmentation via Monte-Carlo Tree Search for Prostate MRI Segmentation}

\author{Xinyue Xu \inst{1}\thanks{Both authors contributed equally to this work.} \and Yuhan Hsi \inst{2}\repeatthanks \and Haonan Wang \inst{3} \and Xiaomeng Li \inst{1}\textsuperscript{(\Letter)}
}

\authorrunning{X. Xu et al.}
\titlerunning{Dynamic Data Augmentation via MCTS for Prostate MRI Segmentation}

% \author{Anonymous}
% \institute{Anonymous Organization}
\institute{The Hong Kong University of Science and Technology, Hong Kong 
\\ \email{xxucb@connect.ust.hk, eexmli@ust.hk}
\and
The Pennsylvania State University, State College PA, USA 
\\ \email{ybh5084@psu.edu}
\and
The University of Hong Kong, Hong Kong 
\\ \email{haonanw@connect.hku.hk}}

%
\maketitle              % typeset the header of the contribution

\begin{abstract}

Medical image data are often limited due to the expensive acquisition and annotation process. Hence, training a deep-learning model with only raw data can easily lead to overfitting. One solution to this problem is to augment the raw data with various transformations, improving the model's ability to generalize to new data. However, manually configuring a generic augmentation combination and parameters for different datasets is non-trivial due to inconsistent acquisition approaches and data distributions. Therefore, automatic data augmentation is proposed to learn favorable augmentation strategies for different datasets while incurring large GPU overhead. To this end, we present a novel method, called Dynamic Data Augmentation (\textbf{DDAug}), which is efficient and has negligible computation cost. Our DDAug develops a hierarchical tree structure to represent various augmentations and utilizes an efficient Monte-Carlo tree searching algorithm to update, prune, and sample the tree. As a result, the augmentation pipeline can be optimized for each dataset automatically. Experiments on multiple Prostate MRI datasets show that our method outperforms the current state-of-the-art data augmentation strategies.

\end{abstract}

\section{Introduction}

The prostate is an important reproductive organ for men. 
The three most prevalent forms of prostate disease are inflammation, benign prostate enlargement, and prostate cancer.
A person may experience one or more of these symptoms. 
Accurate MRI segmentation is crucial for the pathological diagnosis and prognosis of prostate diseases \cite{mahapatra2013prostate}. 
Manual prostate segmentation is a time-consuming task that is subject to inter and intra-observer variability \cite{gardner2015contouring}. 
The development of deep learning has led to significant advancements in many fields, including computer-assisted intervention. 
With the advancement of technology, clinical applications of deep learning techniques have increased. 
There are multiple deep learning networks \cite{milletari2016v, yu2017volumetric, jia2019hd} designed to enhance the accuracy of automatic prostatic segmentation. 
Different neural network structures, such as Vnet \cite{milletari2016v}, U-Net \cite{ronneberger2015u} and its variant nnUNet \cite{isensee2021nnu}, can all be utilized for prostate segmentation. 
These methods are all from the perspective of modifying the model structure to improve the segmentation accuracy. 
However, in medical image segmentation tasks, carefully designed network structure is prone to overfitting due to limited data. 
To alleviate the data shortage, data augmentation is an effective means of enhancing segmentation performance and model generalizability simultaneously on small datasets.

Data augmentation aims to generate more data from the raw samples via pre-defined transformations, which helps to diversify the original dataset~\cite{shorten2019survey}. 
Typical data augmentation techniques include affine transformations (\textit{e.g.}, rotation, flipping, and scaling), pixel-level transformation (\textit{e.g.}, gaussian nosie and contrast adjustment), and elastic distortion. 
For prostate MRI data, affine transformation or GAN-based methods are frequently used \cite{garcea2022data}.
However, the augment selection and combination process that utilizes these aforementioned transformations are predominantly hand-crafted. 
It is difficult to identify which operations are actually useful for the prostate segmentation task, thus often resulting in sub-optimal combinations, or even degrading network performance. 
Automatic data augmentation, with its ability to design different combinations, its flexibility to remove useless augment operations, and its utilization of quantifiable metrics, is a crucial technology that can solve this problem. 
Approaches to automatic data augmentation need to strike a balance between simplicity, cost, and performance \cite{muller2021trivialaugment}. 
In the field of natural image, early automatic augmentation techniques \cite{cubuk2019autoaugment, tian2020improving, lin2019online, ho2019population} were GPU-intensive. 
Subsequently, RandAugment \cite{cubuk2020randaugment}, UniformAugment \cite{lingchen2020uniformaugment}, and TrivialAugment \cite{muller2021trivialaugment} substantially decreased the search cost while maintaining performance. 
Due to the variation between medical (spatial context information and different morphologies of lesions and tissues) and natural images, directly applying these approaches is either ineffective or unsuitable. 
The earliest work \cite{yang2019searching} utilized reinforcement learning (RL) for automatic medical data augmentation, but it required a significant amount of computing resources.
The state-of-the-art automatic data augmentation framework (ASNG) algorithm \cite{xu2020automatic} formulated the automatic data augmentation problem as bi-level optimization and applied the approximation algorithm to solve it. 
Although it is more efficient than reinforcement learning, the time required to find a reasonable strategy can still be highly demanding.
Furthermore, using only rudimentary spatial transforms can limit performance, and some state-of-the-art methods involve searching the probability of operation, which can make the training process inefficient.

To this end, we propose a novel automatic augmentation strategy \emph{Dynamic Data Augmentation ({\methodname})} for MRI segmentation. 
Automatic data augmentation problem is formulated  into the Monte-Carlo tree search problem for the first time.
The augmentation pipeline is represented as a tree structure, which is iteratively refined through updating, pruning, and sampling.
%and the tree is refined iteratively via updating, pruning, and sampling. 
% In contrast to the previous method, we add more spatial augmentations into the search space and allow the tree to determine the appropriate sequence of augmentations while removing useless augmentations. 
In contrast to the previous method, our approach expands the search space by including more spatial augmentations and allows the tree structure to determine the optimal sequence of augmentations while removing redundant ones.
Moreover, our method's flexibility in selecting operations without having to search for the probability significantly enhances its search efficiency. 
% The ability to select operations flexibly without having to search for the probability increases the search's efficiency. 
% Our method only utilizes a few augmentation operations at a time and can achieve the effect of manually combining many operations.
Our method adopts a novel approach by using only a few augmentation operations at a time, yet achieving an effect similar to that of manually combining multiple operations.
Our DDAug method achieves an optimal balance between simplicity, cost, and performance when compared to previous approaches. Code and documentation are available at \href{https://github.com/xmed-lab/DDAug}{https://github.com/xmed-lab/DDAug}.

\section{Methodology}

% We propose an automatic data augmentation strategy that allows zero GPU overhead and is able to outperform the current SOTA approach. We demonstrate the importance of increasing types of operations and the need to search for the correct combination of operations. We also show that simply adding more operations sequentially does not increase performance. Our approach can easily be integrated into any training process to enhance the network performance and robustness on incoming data. 

\begin{figure}[htbp]
    \centering
    \includegraphics[width=0.8\columnwidth]{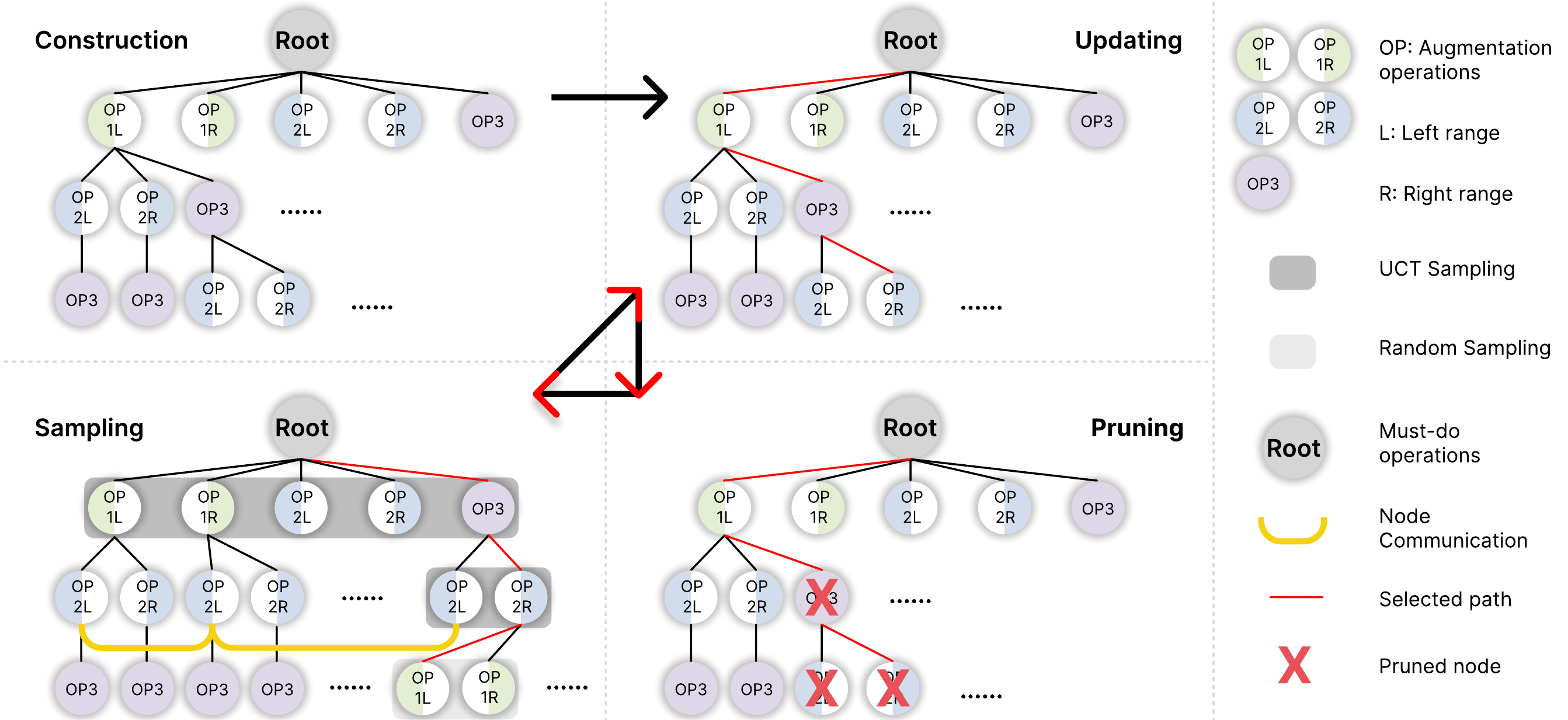}
    \caption{Four Stages of Monte-Carlo Tree Search.}
    % \caption{Four Stages of Monte-Carlo Tree Search. During updating, the selected path is highlighted in red. Use a cross when pruning to remove nodes that hinder performance. During the sampling process, different layers employ distinct sampling mechanisms based on the frequency of visits. The same level of augmentation operation has communications.}
    \label{fig:mcts}
\end{figure}

Automatic augmentation search spaces and Monte-Carlo tree search constitute our method. 
We meticulously selected a number of dependable operations on medical images to compose our search space for the tree's construction. 
The search space consists of pixel-level and spatial-level transformations, as well as the left and right ranges of their respective magnitudes. 
After the tree is constructed, a path is chosen for each training epoch, which is updated by the validation loss, and nodes and their children in the chosen path that degrade model performance are pruned. 
Finally, the path of the subsequent epoch is chosen by random or Upper Confidence Bounds applied to Trees (UCT) \cite{auer2002finite} sampling for different layers, and the cycle continues until training is complete. 
Fig.~\ref{fig:mcts} illustrates the procedure of the complete tree and the whole training process can be summarized in Algorithm~\ref{algo}. We will elaborate on each section below.

% Our method consists of four stages: tree construction, updating, pruning, and sampling. Similar to the standard automatic augment procedure, we have a predefined search space for building our augmentation tree. The search space is comprised of pixel-level and spatial-level transformations, along with their respective magnitude ranges. The tree is constructed methodically so that every path has a unique sequence of augmentation operations, and no two paths have the same order of augmentations. During the training process, every epoch will use exactly one path to obtain the validation loss, and the loss change will be used to update the quality (Q-value) of all the nodes within the path. The updated Q-values are then used to sample a different order of augmentations path for the subsequent epoch. Despite updating node quality, we prune nodes (including their child nodes) that, according to the loss change records, have an overall negative impact on network performance. After pruning, we determine whether to use random sampling or Upper Confidence Bounds applied to Trees (UCT) \cite{auer2002finite} sampling to generate a new path for the next epoch based on the number of times the node is visited. Fig.~\ref{fig:mcts} depicts the entire procedure, and we will elaborate on each section below.

\subsection{Search Space}

The design of our search space is crucial to the performance of the final network. 
To compensate for the absence of spatial-level augmentation operations, optical distortion, elastic transformation, and grid distortion are added. Table~\ref{tab:search_space} displays the complete search space. 
To better assess the efficacy of various operations in relation to their magnitude, we divide the magnitude into left-range and right-range whenever possible.
Operations like brightness transform exhibit left-range (decrease brightness) or right-range (increase brightness).
The random crop operation will pad the data by first padding it with a stochastically selected percentage, then randomly crop the padded data to its original size. 
Unlike brightness transform, random crop only has one magnitude range. 
This division allows for more precise control over the magnitude range without significantly increasing tree size. 
The operations of the root type pertain to the subsequent construct tree.
These are necessary augmentation operations for each path and are not involved in the search. 
We forego the search for augmentation probability for two reasons. 
First, it would significantly increase the size of the tree, making search inefficient. 
Second, if a particular operation and magnitude range combination increases validation performance, it will be sampled more frequently. 
And if the combination is prone to degrade network performance, it will be swiftly removed. 

% We intentionally removed rotation operation d

\begin{table}[htbp]
\centering
\caption{Augmentation Search Space. The root type refers to must-do operations at the beginning of path selection. '-' denotes range not applicable.}
\label{tab:search_space}
\begin{tabular}{c|c|c|cc}
\hline
Operations & LR & RR & Type &  \\ \hline
Mirror           &  -  &  -  &       root      &  \\ \hline
% Rotation & (-30\degree, 30\degree) & -  &      root    &  \\ \hline
Random Crop &  (0\%, 33\%)  &  -  &      root     &  \\ \hline
Contrast Adjustment           &  (0.5, 1)  &  (1, 1.5)  &      pixel-level     &  \\ \hline
Gamma Transform           &  (0.5, 1)  &  (1, 1.5)  &     pixel-level    &  \\ \hline
Brightness Transform           &  (0.5, 1)  &  (1, 1.5)  &      pixel-level     &  \\ \hline
Gaussian Noise           &  (0, 0.1)  &   - &       pixel-level    &  \\ \hline
Gaussian Blur           &  (0.5, 1)  &  (1, 1.5)  &      pixel-level      &  \\ \hline
Simulate low-res image           &  (0.5, 1)  &  -  &     pixel-level       &  \\ \hline
Scale           &   (0.5, 1) &  (1, 1.5)  &        spatial-level   &  \\ \hline
Optical Distortion           &  (0, 0.05)  &  -  &         spatial-level    &  \\ \hline
Elastic Transform           &  (0, 50)  & -   &      spatial-level      &  \\ \hline
Grid Distortion           &  (0, 0.3)  &  -  &       spatial-level    &  \\ \hline
\end{tabular}
\end{table}

\subsection{Tree Construction}

Similar to tree nodes, different augmentation operations can be connected to create various augmentation combinations. 
Additionally, we must consider their order when using a sequence of augmentations. 
This resembles the structure of a tree very closely. 
For efficient search purposes, we encode the automatic data augmentation as a tree structure. 
To construct our tree using the specified search space, we begin by generating a root node with mirror and random crop operations. 
These operations serve as a set of foundational augmentations and are always applied prior to the augmentation path chosen by tree search for each epoch. 
The remaining augmentation operations will participate in the search, and we use a three-layer tree structure to load the search space. 
Each node in the first layer is an augmentation operation, and their child nodes are the augmentation operations that do not include their parent.
There are no duplicate operations on a single path, and no two paths have the same order of operations. 
The first augment path is initialized as the leftmost path of the tree.

% define node height, and method of calculation. introduce and calculate {k_{delete}} and {k_{uct_threshold}}

%After we build the tree, we perform an initial trimming to ensure that no two paths uses a same set of nodes (a node includes both augment operation and magniture range). We start by randomly sampling a path $P_a$ and compare $P_a$ with every other path in the tree. If we discover a path $P_b$ that contains the same set of nodes as $P_a$, we remove the last node from $P_b$. We continue to sample new $P_a$ and compare with the rest of the tree until all nodes have been visited. After the trimming process, we compute the distance of a node from leaf for all nodes in the tree for later use. We initialize the first augment path to be the left most path of the tree for epoch 0. 

\begin{algorithm}[htbp]
 \begin{algorithmic}[1]

\STATE Initialize the augmentation tree.
\STATE Set the leftmost path as the first augment path. 

\FOR{each epoch}
\STATE Train the model and calculate the validation loss.
\FOR{each node in previously selected path}
    \STATE Update node Q-value (Eq.~\ref{eq:2}) using moving average loss.
    \STATE Record validation loss change $L_{node}$.
    \IF{Past $\sum_{n=1}^{5} L_{node} >$ 0}
        \STATE Delete the current node and subtree of this node.
        \STATE Break; 
    \ENDIF
\ENDFOR
\WHILE{not at leaf node}
    \IF{mean visited times $<$ $k_{uct}$}
        \STATE Sample node using Random sampling. 
    \ELSE
        \STATE Sample node using UCT sampling (Eq.~\ref{eq:5}). 
    \ENDIF
\ENDWHILE
\STATE Finish sampling path for next epoch. 
\ENDFOR

\STATE Inference and report testing data performance.

\caption{Training Process of \methodname}
\label{algo}
\end{algorithmic}
\end{algorithm}

\subsection{Tree Updating and Pruning}

% Since we sample exactly one path for every epoch, there are no changes made to the tree until a new validation loss is computed at the end of an epoch. 
With the initialized path during tree construction, we train the model for one epoch per path and calculate the validation loss $L_{val}$. The validation loss is computed utilizing the original nnUNet CE + DICE loss. The validation loss is then employed to update the tree by calculating the moving average loss $L_{ma}$ using the following formula:
% add loss is CE + (1-DICE)

\begin{equation} \label{eq:1}
    L_{ma} = \beta \cdot L_{val}^{t-1} + (1 - \beta) \cdot L_{val}^{t}
\end{equation}

where $\beta \in [0, 1]$ controls the ratio of current validation loss. $L_{val}^{t-1}$ is the validation loss of the previous epoch, while $L_{val}^{t}$ represents the validation loss of the current epoch. We then update the Q-value of all nodes in the previously selected path with: 

\begin{equation} \label{eq:2}
    Q = \frac{L_{ma}}{L_{val}^{t}}
\end{equation}

A record of validation loss change $L_{node} = L_{val}^{t} - L_{val}^{t-1}$ 
% \begin{equation} \label{eq:3}
%     L_{node} = L_{val}^{t} - L_{val}^{t-1}
% \end{equation} 
is kept for all nodes to evaluate their overall impact on network performance. As we traverse the path to update the Q-value, if the sum of the previous five $L_{node}$ scores is greater than 0, the node is deemed to have a negative impact on the network's performance and is pruned from the tree.

\subsection{Tree Sampling}

After the pruning process, a new path needs to be chosen for the subsequent epoch. 
Because the nodes have not yet been visited, we use random sampling rather than Monte-Carlo UCT sampling at the beginning of the network training process. 
We compare the $k_{uct}$ threshold to the average visited times of the current layer to determine when to switch to UCT sampling. 
The value of $k_{uct}$ is set to 3, 1, and 1 times, for the first, second, and third layers of the tree, respectively. 
The number of tree layers is expandable, but increasing the number of layers will lead to the exponential growth of search space, which is not conducive to search efficiency. 
% At the same time, if the tree is less than three layers, all nodes are easily visited, and it is difficult for the scoring mechanism to give reasonable feedback. So we chose to use three layers as the final number of layers.
At the same time, if the tree has less than three layers, the amount of nodes and paths is extremely limited, thus decreasing the diversity introduced via data augmentation. 

Inspired by \cite{su2021prioritized}, we introduce a node communication term $S$ to better evaluate the current node's efficacy using nodes' Q-value of nodes from the same layer that has the same augment operation and magnitude range as the current node. 

\begin{equation} \label{eq:4}
    S(v_i^{l}) = (1 - \lambda) \cdot Q(v_i^{l}) + \lambda \cdot \sum_{j=0}^{n} \frac{Q(v_j^{l})}{n} 
\end{equation}

where $v_i^{l}$ is the $i$-th child node in the $l$-th layer, $v_j^{l}$ is the other nodes that have the same operation and magnitude range as $v_i$ in the $l$-th layer, and $n$ denotes the total number of $v_j^{l}$. 
$\lambda$ controls the effect of the node communication term. 

% add node communication 
When the averaged visited times of all children of the current node exceeds $k_{uct}$, we employ the following equation to calculate the UCT \cite{kocsis2006bandit, su2021prioritized} score for all children of the current node:
\begin{equation} \label{eq:5}
    UCT(v_i^{l}) = \frac{Q(v_i^{l})}{n_i^{l}} + C_1 \sqrt{\frac{log(n_p^{(l-1)})}{n_i^{l}}} + C_2 \cdot S(v_i^{l})
\end{equation}
%left is exploitation, right is exploration 
where $n_i^{l}$ is the number of visited times of $v_i^{l}$, and $n_p^{(l-1)}$ is the visited times of its parent node in the $(l-1)$-th layer. 

A temperature term $\tau$ \cite{neumann2018relaxed} is utilized to promote greater discrimination between candidates by amplifying score differences, thus the sampling probability can be calculated as 

\begin{equation} \label{eq:6}
    P(v_i^{l}) = \frac{\exp( \frac{UCT(v_i^{l})}{\tau} )}
    {\sum_j^{n} \exp(\frac{UCT(v_j^{l})}{\tau})}
\end{equation}

where $v_i^{l}$ are children of the current node. We sample a node from the current group of children using the probabilities calculated, then continue the sampling process until a leaf node is reached. 
Reaching a leaf node signifies the termination of the sampling process for the current epoch, and the selected path will be adopted in the next epoch. 
This cycle repeats at the end of every epoch until maximum the training epochs are reached. 

% Once a leaf node is reached, the sampling process terminates. 
% The newly selected path will be used during the next epoch. 
% The sampling process repeats at the end of every epoch after validation loss is acquired.

\section{Implementation and Experiments}

\subsection{Datasets and Implementation Details}

% nnUNet Task05\_Prostate \cite{simpson2019large} manual segmentation of the whole prostate from transverse T2-weighted scans.

% Multi-site Dataset for Prostate MRI Segmentation \cite{liu2020ms}

\subsubsection{Datasets.}
We conduct our experiments on several 3d Prostate MRI datasets: subset 1 and 2 are from NCI-ISBI 2013 challenge \cite{prostate_subsetAB}, subset 3 is from I2CVB benchmarking \cite{LEMAITRE20158}, subset 4, 5, 6 are from PROMISE12  \cite{prostate_subsetdef}, and subset 7 is the Task 005 prostate dataset from Medical Segmentation Decathlon \cite{simpson2019large}. 
% and Multi-site Dataset for Prostate MRI Segmentation \cite{liu2020ms}. 
% briefly describe dataset 
Subsets 1 through 6 are acquired from and have been resized by \cite{liu2020ms}. 
All datasets are then resampled and normalized to zero mean and unit variance as described in nnUNet \cite{isensee2021nnu}.

% We did not reproduce ASNG for all the experiments due to its prohibitaly long training time. 

% moreDA -- scaling, rotation, gaussianNoise, gaussianBlur, BirghtnessMultiplicative, contrast augmentation, simulate low resolution, gamma transform, mirror  

% ASNG nnUNet 0.6 gpu hours
% describe how experiments are conducted. explain experiment setting. we run nnunet with noaug etc, moreda, mcts.... 
% didnt reproduce ASNG b/c nnunetV1, gpu hrs 

\subsubsection{Implementation Details.}
For a fair comparison, we base our implementation on the original nnUNet repository. 
We only inserted additional code for the implementation of DDAug while keeping model architecture and self-configuration process intact. 
To conduct 5-fold cross-validation, we utilized stochastic gradient descent with Nesterov momentum and set the learning rate of 0.01. 
Each fold trains for 200 epochs, and each epoch has 250 batches with a batch size of 2. 
% Given that there are no additional GPU costs, training time with DDAug was comparable to training time with nnUNet, which is approximately 30 GPU hours on a single NVIDIA GeForce RTX 3090.
% The speed demonstrates the efficiency of our tree search. Our code will be available after acceptance. 
The runtime comparison can be found in Table \ref{tab:gpu}. The utilization of Reinforcement Learning and ASNG method demand substantial GPU resources. In contrast, our approach performs at an equivalent efficiency to the original nnUnet data augmentation.

\begin{table}[htbp]
\centering
  \caption{Comparison of GPU costs with different augmentation methods.}
  \begin{tabular}{c|c|c|c|cl}
    \hline
    Method & RL \cite{yang2019searching} & ASNG \cite{xu2020automatic, he2022differentiable} & DDAug \textbf{(Ours)}  & nnUNet \cite{isensee2021nnu}\\
    \hline
    Cost (hours) & 768 & 100 & 40 & 40  \\
  \hline
\end{tabular}
\label{tab:gpu}
\end{table}

\subsubsection{Compared Methods.}
% Since the previous version of nnUNet used by ASNG~\cite{xu2020automatic} requires ten days of GPU processing time and our objective is to create an effective and efficient automated search method, we only compare these to the latest version of nnUNet.
Since ASNG~\cite{xu2020automatic} requires ten days of GPU processing time and our objective is to create an effective and efficient automated search method, we only compare our implementations on nnUNet that has the same GPU runtime requirements.
Limited by the size of each subset, we conduct all of our experiments using 5-fold cross-validation and report the mean validation DICE score inferred with weights from the last epoch. 
Our baselines are established via training nnUNet using no augmentations (NoDA) and using default augmentations (moreDA). 
The `moreDA' is a set of sequential augmentations including scaling, rotating, adding gaussian noise, adding gaussian blur, transforming with multiplicative brightness, transforming with contrast augmentation, simulating low resolution, performing gamma transform, and mirroring axis.

\subsubsection{Ablation Study.}

In our ablation study, we start by replacing the sequential augment operations in moreDA with the uniform sampling mechanism described TrivialAugment \cite{muller2021trivialaugment}. 
This allows us to assess the viability of using the natural image SOTA approach on medical image.
% TrivialAugment replace the sequential augment operations in moreDA with uniform sampling of one operation while retaining the augment search space.
To evaluate the effectiveness of the proposed search space, we extend moreDA's operation search space with additional spatial augmentations (Spatial SS). 
Finally, we replace moreDA with DDAug to examine the advantage of using the expanded search space inconjunction with Monte-Carlo Tree Search (MCTS). 

% \begin{table}[htbp]
% \centering
%   \caption{Comparison of GPU costs.}
%   \begin{tabular}{c|c|c|c|cl}
%     \hline
%     Method & RL \cite{yang2019searching} & ASNG \cite{xu2020automatic} & DDAug \textbf{(Ours)}  & nnUNet \cite{isensee2021nnu}\\
%     \hline
%     Cost (hours) & 768 & 100 & 30 & 30  \\
%   \hline
% \end{tabular}
% \label{tab:gpu}
% \end{table}

% \COMMENT{
% nvidia 3090, etc, 200epoch, 250, etc. change from nnunet etc. 
% refer to list of question filled during MICCAI submission
% }

\subsection{Experimental Results}

\begin{table}[htbp]
\centering
\caption{Augmentation performance of different Prostate datasets on Dice (\%). Subsets represent different prostate datasets and the backbone is nnUNet. NoDA: No augmentation; moreDA: sequential augmentation; Spatial SS: our designed search space; TrivialAugment: natural image SOTA method; 
DDAug: MCTS + our search space (proposed method). 
\textcolor{red}{red}, \textcolor{blue}{blue} denote the highest and second highest score.}
\resizebox{\linewidth}{!}{\begin{tabular}{c||c|c|c|c|c|c|c||c}
\hline
Method       & Subset 1 & Subset 2 & Subset 3 & Subset 4 & Subset 5 & Subset 6 & Subset 7  &  Average\\ \hline
NoDA &   79.12     &   80.82     &   84.57     &    82.02    &   78.10     &   82.77   &  72.36 &  79.97 \\ \hline
moreDA       &   79.64   &    81.66    &     87.60   &    81.38    &    83.74    &    \textcolor{blue}{87.12}  &  71.23  &    81.77  \\ \hline
% TrivialAugment   &    80.39    &  82.30    &    \underline{88.83} &    84.01    &       86.60    &     86.12   &    -  \\ \hline

TrivialAugment & \textcolor{red}{80.39} & \textcolor{blue}{82.21} & \textcolor{red}{88.42} & 82.60 &  \textcolor{blue}{86.36} & 86.60 & 72.58 &  \textcolor{blue}{82.74}\\ \hline 

Spatial SS \textbf{(Ours)}      &   79.96   &   82.18     &   \textcolor{blue}{87.68}  &   \textcolor{blue}{83.74}    &   85.69 &   86.99  & \textcolor{blue}{72.90}   &    82.73  \\ \hline

% Spatial SS Uniform \textbf{(Ours)}      &   79.91   &   82.28     &   87.26  &   89.77    &   86.09 &   85.60  & 72.16   &    83.29  \\ \hline

DDAug \textbf{(Ours)}      &    \textcolor{blue}{80.27}    &   \textcolor{red}{82.72}     &     87.46   &    \textcolor{red}{88.59}    &    \textcolor{red}{86.40}    &   \textcolor{red}{87.17}   & \textcolor{red}{73.20}     &  \textcolor{red}{83.69}  \\ \hline

\end{tabular}}
\label{tab:results}
\end{table}

The five-fold average Dice similarity coefficients of different methods are shown in Table~\ref{tab:results}. 
As we can see, in general, adding augmentation to the prostate MRI dataset is better than no augmentation. 
moreDA demonstrates some improvement from NoDA on most of the datasets, and additional performance increase are observed when expanding the search space by adding spatial-level augmentations. 
When comparing the performance of Spatial SS and TrivialAugment, the improvement prove to be inconclusive, as three out of seven dataset exhibits degradation. 
This is likely due to the fact that TrivialAugment uses uniform sampling over the search space, and unlike our DDAug, does not consider the efficacy of different operations. 
% TrivialAugment also suffers from less diversed augmented data since it only uses one augmentation at a time. 
We are able to further improve the results by utilizing DDAug's full search space and its tree search method. 
It is important to note that moreDA contains 9 sequential augmentations while DDAug only uses 5. This indicates that simply piling on more augmentations sequentially is not the optimal solution. 
Though using only a few operations per epoch, DDAug still achieves the highest average DICE when looking at all 7 subsets with near-zero computing consumption. 
% Finally, we achieved the best results both on the average multi-site subsets and the original nnUNet prostate datasets. 
% We also compare with the SOTA method in the natural image to validate the effectiveness of our DDAug and the results in Table~\ref{tab:ab} demonstrates our method outperforms the trivial augmentation.

\begin{figure}[htbp]
    \centering
    \includegraphics[width=1.0\columnwidth]{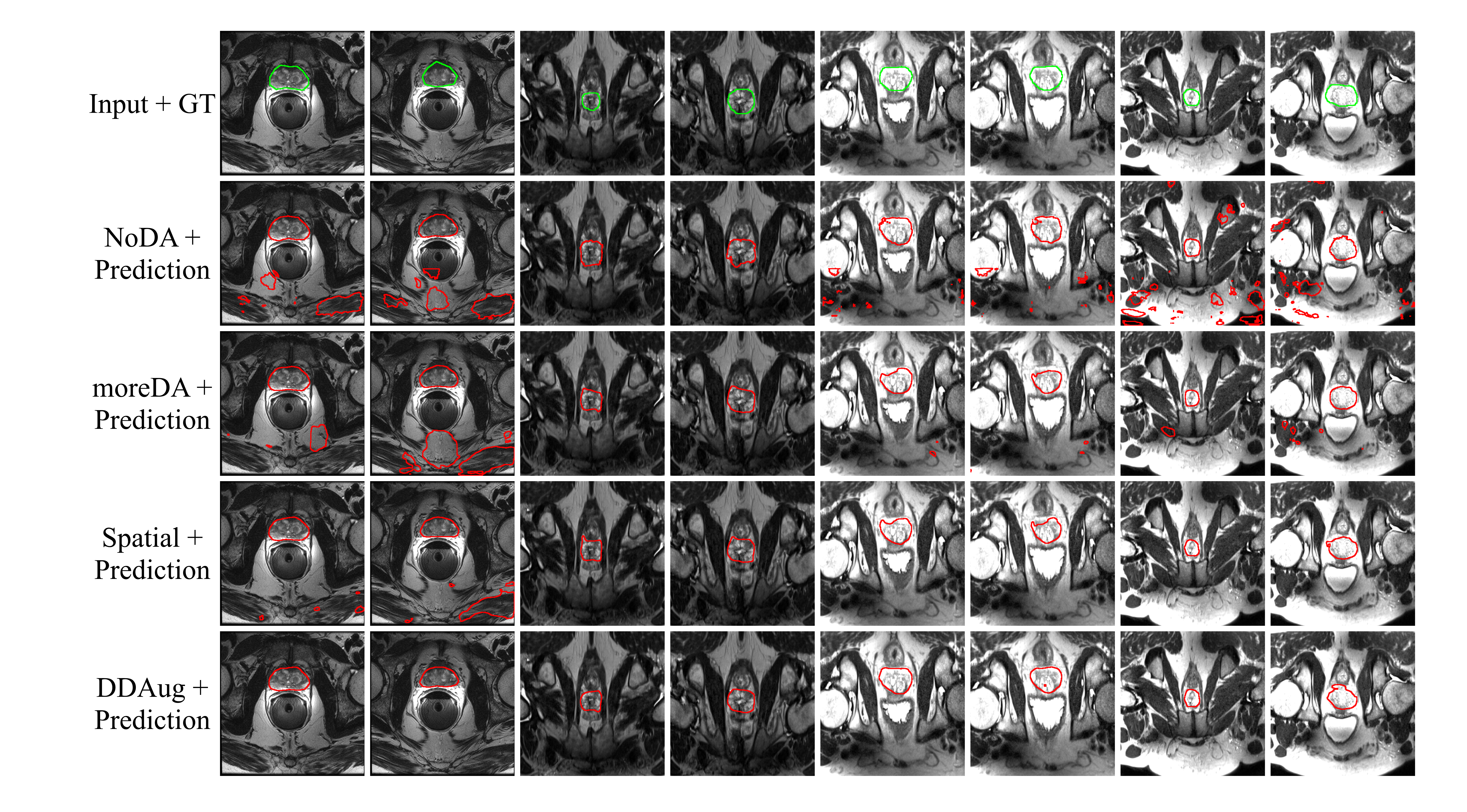}
    \caption{Comparison of inference results using different augmentation techniques during training. The top row is validation images and their corresponding ground truth. The subsequent rows are inference results using models trained with no augmentation, moreDA augmentation, our designed search space, and DDAug, respectively. }
    \label{fig:ar}
\end{figure}

The performance difference can translate to visual discrepancy between different methods. 
When inspecting validation segmentation results, we noticed that DDAug is significantly more robust when segmenting validation cases as shown in Fig. \ref{fig:ar}. 
DDAug demonstrates enhanced generalizability against its counterparts. 
Augmenting data sequentially, on the other hand, was not able to handle difficult validation cases. 

% \begin{table}[htbp]
% \centering
% \caption{Ablation study compared with Trivial Augmentation (Dice).}
% \begin{tabular}{c|c|c|c}
% \hline
% Subset  & moreDA & TrivialAugment &  DDAug          \\ \hline
% Subset2 & 81.66  & 82.30  & \textbf{82.72} \\ \hline
% Subset4 & 81.38  & 84.01  & \textbf{88.59} \\ \hline
% Subset6 & 87.12  & 86.12  & \textbf{87.17} \\ \hline
% \end{tabular}
% \label{tab:ab}
% \end{table}

\section{Conclusion}

We propose an efficient and zero GPU overhead automatic data augmentation algorithm for prostate MRI segmentation. Comparing previous approaches, we include additional spatial transformations into the search space, and adopt a Monte-Carlo tree structure to store various augmentation operations. An optimal augmentation strategy can be obtained by updating, pruning, and sampling the tree. Our method outperforms the state-of-the-art manual and natural image automatic augmentation methods on several prostate datasets. We show the feasibility of utilizing automatic data augmentation without increasing GPU consumption. In future work, we will further investigate the generalizability of tree search on other medical segmentation datasets, \textit{e.g.}, liver cancer segmentation, brain tumor segmentation and abdominal multi-organ segmentation. 

% We propose an efficient and GPU additional-cost-free automatic data augmentation algorithm for prostate MRI segmentation. 
% The conventional search space is added with spatial transformations, and the Monte-Carlo tree structure is then used to store various augmentation operations. 
% By updating, pruning, and sampling the tree, a reasonable augmentation strategy is obtained. 
% Our method outperforms the state-of-the-art manual and natural image automatic augment methods across several prostate datasets. 
% The proposed method demonstrates that effective data augmentation is possible via automatic search without increasing GPU consumption. 
% % In future work, we will examine the universality of tree search by applying to data of additional modalities.
% In future work, we will examine the generalizability of tree search by incorporating data from other tasks such as liver cancer segmentation, brain tumor segmentation, abdominal multi-organ segmentation, etc. 

% \begin{figure}[htbp]
%     \centering
%     \includegraphics[width=0.8\columnwidth]{Frame.pdf}
%     \caption{Visualization of Augmentation Results.}
%     \label{fig:ar}
% \end{figure}

\section{Acknowledgement}
This work was supported by the Hong Kong Innovation and Technology Fund under Project ITS/030/21, as well as by Foshan HKUST Projects under Grants FSUST21-HKUST10E and FSUST21- HKUST11E.

\bibliographystyle{splncs04}
\bibliography{bib}




% LNCS Homepage, \url{http://www.springer.com/lncs}. Last accessed 4
% \end{thebibliography}
\end{document}